# Synchronized resistance to inhomogeneous magnetic field-induced dephasing of an image stored in a cold atomic ensemble

Ying-Hao Ye[1,2], Lei Zeng[1,2], Yi-Chen Yu[1,2], Ming-Xin Dong[1,2], En-Ze Li[1,2], Wei-Hang Zhang [1,2], Zong-Kai Liu[1,2], Li-Hua Zhang[1,2], Guang-Can Guo[1,2], Dong-Sheng Ding[1,2,§], and Bao-Sen Shi [1,2,*]

**Long-lived storage of arbitrary transverse multimodes is important for establishing a high-channel-capacity quantum network. Most of the pioneering works focused on atomic diffusion as the dominant impact on the retrieved pattern in an atom-based memory. In this work, we demonstrate that the unsynchronized Larmor precession of atoms in the inhomogeneous magnetic field dominates the distortion of the pattern stored in a cold-atom-based memory. We find that this distortion effect can be eliminated by applying a strong uniform polarization magnetic field. By preparing atoms in magnetically insensitive states, the destructive interference between different spin-wave components is diminished, and the stored localized patterns are synchronized further in a single spin-wave component; then, an obvious enhancement in preserving patterns for a long time is obtained. The reported results are very promising for studying transverse multimode decoherence in storage and high-dimensional quantum networks in the future.**

[1]Key Laboratory of Quantum Information

[2]Synergetic Innovation Center of Quantum Information and Quantum Physics

[§]dds@ustc.edu.cn.　[*]drshi@ustc.edu.cn

# INTRODUCTION

As a robust carrier of quantum information, a photon has many degrees of freedom, such as frequency, polarization, and transverse multimode[1,2], in which information can be encoded. Among them, the transverse multimode has received great attention because transverse modes such as Laguerre-Gaussian modes[3-6] can form an infinite-dimensional Hilbert space. By encoding information in transverse multimodes, one can dramatically increase the channel capacity of quantum information processing[7,8]. In the quantum information field, due to the long coherence time between metastable states of atoms, a long-lived quantum memory based on an atomic ensemble[9-11] can be achieved, which is promising for realizing quantum repeaters to overcome the strong attenuation of optical channels in a long-distance quantum network[12-20]. A quantum memory that can store quantum states for a long time can increase the success probability for entanglement creation per round-trip time and thereby decrease the time to establish entanglement between end nodes in the quantum network[16]. Thus, studying the long-lived storage of spatial transverse multimodes in atomic ensembles is increasingly demanded[2,8,21-26] for establishing a high-capacity quantum network.

Many significant works on storing transverse multimodes have been realized in a cold atomic ensemble[21,27,28]. however, the achieved time for storing an image was only up to several microseconds. Although many works have achieved long-term storage in atomic ensembles[9-11], the stored field is still a single-mode field. Since the pioneering work of *Pugatch*[29] proved that optical modes with phase singularities are robust to strong diffusion of hot atoms in an atomic vapor, many elaborate schemes have been proposed to store transverse multimodes for a long time[30]. Techniques such as utilizing the diffraction property of lenses[31,32], employing coupling light with a tailored transverse phase analogous to phase-shift lithography[33] or making use of the ghost imaging technique[34] have been used to realize image storage up to tens of microseconds. However, when we consider transverse multimode storage in a cold atomic ensemble, the dominant deteriorating effect originating from atomic motion considered in previous works can be neglected during short storage time on the scale of microseconds; therefore, exploring other dephasing mechanism in storing transverse multimodes becomes significantly important.

Here, we find that the inhomogeneous magnetic field-induced dephasing effect is the dominant source of the distortion of stored patterns. Usually, a complex pattern has more complicated spectrum distributions on the Fourier plane (FP) compared to the Gaussian mode; therefore, the spatial overlap between the diffracted pattern and ambient magnetic field is larger, and the stored pattern is thus more vulnerable to the unsynchronized evolution of the stored spectrum caused by the magnetic field gradient. In this work, we describe this distortion effect based on the dark state polariton (DSP) evolution theory[35,36]. According to the theory, we propose a method to prolong the storage time of transverse multimodes by applying a strong uniform polarization magnetic field that can synchronize the Larmor precession of atomic magnetic moments at different localized positions. The experimental observations are in good agreement with the simulated results. To diminish destructive interference between different spin-wave components, we prepare the atomic ensemble in magnetically insensitive states, and finally, the storage time of transverse multimodes is prolonged by two orders of magnitude. Our method requires neither establishment of a sophisticated system to compensate for the inhomogeneous field[37] nor the use of a magnetic shield made of a high-permeability material[38].

**EXPERIMENTAL SETUP AND MODEL**

The experiments were performed within the D1 transition of rubidium 85 (Rb 85), as depicted in Fig. 1 (a). We denote $|5S_{1/2}, F=2\rangle$, $|5S_{1/2}, F=3\rangle$ and $|5P_{1/2}, F=3\rangle$ as $|g\rangle$, $|s\rangle$ and $|e\rangle$, respectively. The coupling light is locked on the $|s\rangle \leftrightarrow |e\rangle$ transition, and the centre frequency of the signal field is resonant with the $|g\rangle \leftrightarrow |e\rangle$ transition.

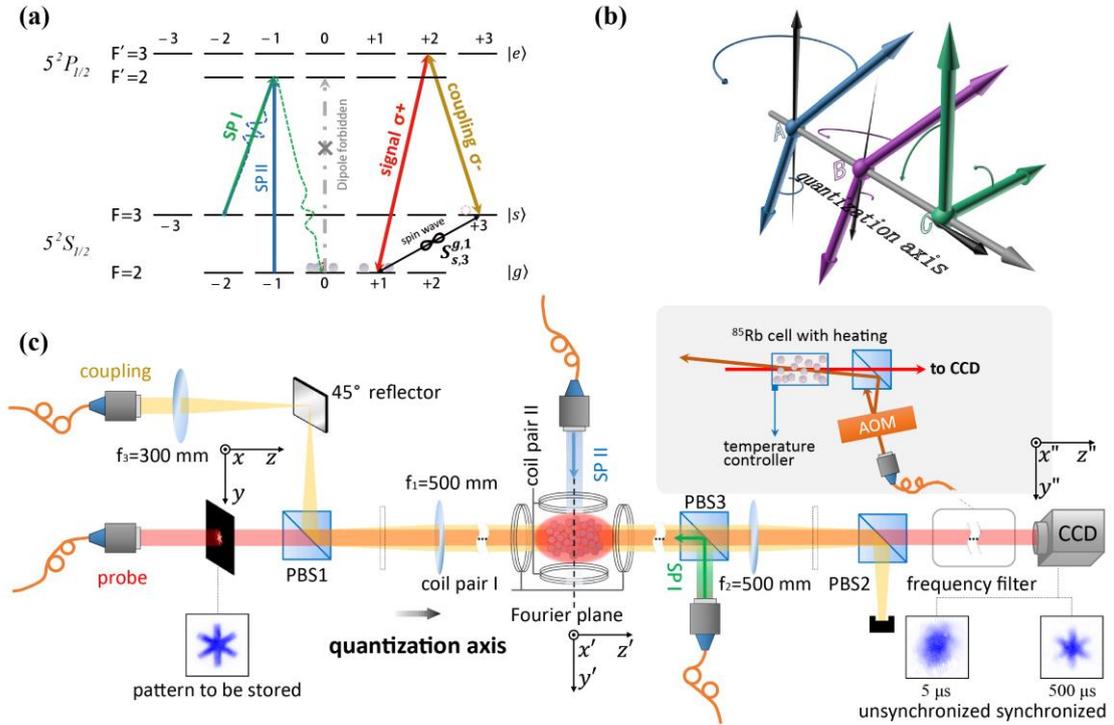

**Fig. 1 Experimental setup and relative energy levels.** (a) Energy levels used in the experiment. When the signal light and coupling light are orthogonally circularly polarized, four groups of spin waves exist in an unpolarized ensemble. After state preparation (SP), only one group of spin waves exists. (b) Schematic of the Larmor precession of atomic magnetic moments. The precession behaviour of two initially identical spin waves is quite different in inhomogeneous magnetic fields, leading to different phases and intensities of DSPs after precession. At point C, when the polarized magnetic field is parallel to the quantization axis and the direction of the atomic magnetic momentum is perpendicular to the quantization axis, the DSPs undergo the same evolution during precession. (c) Experimental setup. PBS, polarization beam splitter; QWP, quarter-wave plate; AOM, acoustic-optical modulator; CCD, ICCD camera.

Our experimental arrangement is illustrated in Fig. 1 (c). An atomic ensemble of Rb85 is trapped in a three-dimensional magneto-optical trap (MOT). Three pairs of mutually orthogonal Helmholtz coils are placed around the MOT to compensate for the static uniform magnetic field near the ensemble. Strong coupling light with vertical polarization and weak signal light with horizontal polarization are input into the ensemble. To diminish decoherence from atomic motion, we let the signal light and coupling light propagate collinearly after PBS I to suppress the wavenumber of the spin wave. The signal light passes through a pattern mask, and the coupling light is focused by a lens with a focal length of $f_3$=300 *mm*. The mask is placed at the front focal plane of the lens before the MOT with a focal length of $f_1$=500 *mm*, and the spatial centre of the MOT coincides with the back focal plane. After passing through lens $f_1$, the coupling light is shaped as a Gaussian beam with a waist diameter of 4.2 mm. Lens $f_1$ also works as a Fourier transformer for the signal light and produces Fraunhofer diffraction of the pattern on the mask at the FP. For arbitrary patterns on the mask, the diffracted patterns in the atomic ensemble are robust to atomic diffusion due to the phase flipping between adjacent bright and dark spots[29, 31, 32]. The lens after the MOT with a focal length of $f_2$=500 *mm*, together with lens $f_1$, forms a 4*f* imaging system for the signal light; thus, the pattern of the mask is imaged on the ICCD (1,024×1,024, iStar 334T series, Andor).

The applied inhomogeneous magnetic field is from a pair of anti-Helmholtz coils (coil pair II in Fig. 1 (c)). The strength and structure of this inhomogeneous magnetic field can be calculated from the measured parameters of coil pair II by using the Biot-Savart-Laplace law (see the Appendix section). When the frequency of the signal light is scanned by an AOM, an EIT window appears at the two-photon resonance $\Delta\omega=\omega_s-\omega_c=\omega_{sg}$[39] (see the Appendix section), where $\omega_s$ and $\omega_c$ are the frequencies of the signal and coupling light, respectively, and $\omega_{sg}$ is the energy difference between |*s*> and |*g*>. We imprint a signal pulse that is resonant with the |*g*>↔|*e*> transition onto the atomic ensemble, and then, this light pulse is transformed into DSPs[40, 41]. By adiabatically turning off the coupling light, the group velocity of DSPs decreases to zero, and the photon components of DSPs are mapped into collective Raman coherences between |*s*> and |*g*>[42]. The Raman coherence can be seen as a spin wave, and we denote the spin wave between |*g, m₁*> and |*s, m₂*> as $S_{s,m_2}^{g,m_1}$. When the signal light and coupling light are orthogonally circularly polarized, four groups of spin waves exist. As illustrated in Fig. 1 (b), the magnetic moments of atoms in |*g, m₁*> and |*s, m₂*> precess around

the magnetic field; hence, the phase and population of different groups of spin waves change, which results in the collapse and revival of DSPs and then causes enhancement and reduction of the retrieval efficiency over the storage time. Because the beam waist of the expanded Gaussian coupling light is large compared to the diameter of the input pattern (~1.6 mm), we approximate it as a uniform plane wave hereafter. Therefore, both the intensity and phase information of the signal field in the FP are continuously converted into the spin wave $\sigma_{gs} = -gE/\Omega_c(t)$, where $E$ is the signal field and $\Omega_c(t)$ is the Rabi frequency of the coupling light[40].

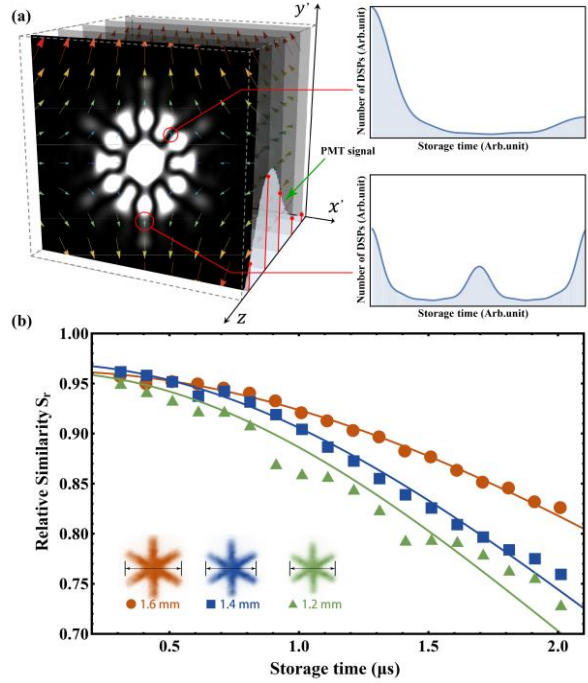

**Fig. 2 Theoretic simulations. (a)** The localized inhomogeneous magnetic field at different positions leads to different evolution behaviours of spin waves; thus, the deformation of the intensity and phase distribution of DSPs with time can be calculated. We use the weighted average of the data along the z-axis as the simulation result. **(b)** Relative similarity as a function of time for patterns of different sizes. The solid line is guided for eye.

To simplify the theoretical analysis, we treat both the signal field and coupling light as monochromatic fields and assume that atoms are initially unpolarized. We establish the Cartesian coordinate system shown in Fig. 2, in which the quantization axis z is parallel to the propagation direction of the signal field. The signal light propagates as DSPs after entering the atomic ensemble, and the annihilation operator of DSPs has the following form[35]:

$$\hat{\Psi}(x',y',z,t) = \frac{i\Omega(t)\hat{a}_\alpha - \sqrt{Np}\kappa^* \sum_m R_m(\alpha,\beta)\hat{S}^{g,m}_{s,m+\alpha-\beta}(x',y',z,t)}{\sqrt{\Omega^2 + Np|\kappa|^2 \sum_m R_m^2(\alpha,\beta)}} \quad (1)$$

where $\alpha$ and $\beta$ are the helicities of the polarization of the signal light and coupling light, respectively, $\hat{a}_\alpha$ is the annihilation operator of the signal photon, $N$ is the number of atoms in the ensemble, $\kappa$ is the coupling constant for the signal transition, $p = 1/(2F_g + 1)$, and the real coefficient $R_m(\alpha,\beta) = C^{F_g,1,F_e}_{m,\alpha,m+\alpha} / C^{F_s,1,F_e}_{m+\alpha-\beta,\beta,m+\alpha}$ is determined by CG coefficients related to the spin waves. After the coupling light is completely turned off, the photonic components of DSPs transform into spin-wave components, and then, the annihilation operator of DSPs has the following form:

$$\begin{aligned}\hat{\Psi}(x',y',z,t_s=0) &\propto \sum_m R_m(\alpha,\beta)\Gamma(x',y',z)\hat{S}^{g,m}_{s,m+\alpha-\beta}(x',y',z,t_s=0) \\ &\propto \sqrt{P(z)}\sum_m R_m(\alpha,\beta)\Gamma(x',y',z)u_f(x',y',t_s=0)\end{aligned} \quad (2)$$

where $u_f(x',y',t_s=0)$ is the amplitude of the signal light in the FP. $P(z)$ is a phenomenological coefficient that describes the weight of the spin wave at different $z$ and satisfies $\int_{-\infty}^{+\infty} P(z)dz = 1$. $\Gamma(x',y',z)$ is the distribution function taking into account the spatial variation of the atomic ensemble density. For a spherical-shaped atomic ensemble with a Gaussian density distribution, we have $\Gamma(x',y',z) \propto exp[-(x'^2 + y'^2 + z^2)/\sigma^2]$, where $\sigma$ is the measured radius of the atomic ensemble, and $u_f$ can be calculated by

$$u_f(x',y',t_s=0) \propto \frac{1}{i\lambda f_1}\iint_\infty u_o(x,y,t_s=0)exp\left[-i\frac{2\pi}{\lambda f_1}(x'x + y'y)\right]dxdy \quad (3)$$

where $u_o$ is the amplitude of the signal field in the object plane of the 4$f$ system and is proportional to the square root of the intensity information recorded by the ICCD. We assume that $u_o$ has a uniform phase distribution because the pattern mask is illuminated by collimated signal light. The evolution of the retrieved signal with time depends on both the intensity and orientation of the ambient magnetic field with respect to the signal vector. Therefore, the evolution of DSPs at different points in the transverse plane is different due to the gradient of the inhomogeneous magnetic field:

$$\hat{\Psi}(x', y', z, t_s = t)$$
$$\propto \sum_{m'} \sum_{m} R_{m'} R_m \left[ D_{m',m}^g \right]^\dagger D_{m'+\alpha-\beta, m+\alpha-\beta}^s \hat{S}_{s,m+\alpha-\beta}^{g,m}(x', y', z, t_s = 0) \quad (4)$$
$$\propto \sqrt{P(z)} \sum_{m'} \sum_{m} R_{m'} R_m \left[ D_{m',m}^g \right]^\dagger D_{m'+\alpha-\beta, m+\alpha-\beta}^s \Gamma(x', y', z) u_f(x', y', t_s = 0)$$

The evolution difference between DSPs at different points in the gradient magnetic field results in deformation of the intensity and phase information in the FP and thus leads to downgrading of the similarity. In a stationary ambient magnetic field $\partial_t \mathbf{B} = 0$, the matrix element of the rotation operator between different quantized levels of hyperfine level F is $D_{m',m}^F(x', y', z, t) = \langle F, m' | exp(-i g_F \mu_B \mathbf{B}(x', y', z) \cdot \hat{\mathbf{F}} t / \hbar) | F, m \rangle$, where $\hat{\mathbf{F}}$ is the total angular momentum operator, $g_F$ is the Landé g factor of hyperfine level F and $\mu_B$ is the Bohr magneton. The quantization axis is chosen to be parallel to the signal field direction. The retrieved signal field intensity is proportional to the number of DSPs. After the second coupling pulse illuminates the ensemble, we have the retrieved signal $|u_f(x', y', t_s = 0)|^2 = \langle \hat{\Psi}^\dagger \hat{\Psi} \rangle$. We denote the calculated relative retrieval efficiency at each transverse point by $\eta$ and the corresponding additional phase angle at that point by $\phi$; taking the quasi-bosonic property of the spin wave into consideration,

$$\eta(x', y', z, t) \propto |\Gamma(x', y', z)|^2 \left| \sum_{m',m} R_{m'} R_m \left[ D_{m',m}^g \right]^\dagger D_{m'+\alpha-\beta, m+\alpha-\beta}^s \right|^2 \quad (5)$$
$$\phi = Arg\left( \sum_{m',m} R_{m'} R_m \left[ D_{m',m}^g \right]^\dagger D_{m'+\alpha-\beta, m+\alpha-\beta}^s \right)$$

Because $P(z)$ satisfies the normalization condition, the retrieved pattern has the following form:

$$I(x'', y'', t_S = t) \propto \int_{-r_a/2}^{r_a/2} dz P(z) \mathfrak{F}\left[ e^{-i\phi} \sqrt{\eta(x', y', z, t)} u_f(x', y', t_s = 0) \right] \quad (6)$$

To numerically simulate the retrieved signal, we need to sample Eq. (6). Because $P(z)$ does not have a singular point, the normalization property can be approximated as $\sum_{z'=-R}^{R} P(z') \delta z' = 1$, where R is the radius of the atomic ensemble, and we have

$$I(x'', y'', t_S = t) \propto \sum_{z'} P(z') \mathfrak{F}\left[ e^{-i\phi} \sqrt{\eta(x', y', z', t)} u_f(x', y', t_s = 0) \right] \delta z' \quad (7)$$

The similarity between the retrieved and original images is quantitatively defined by:

$S = \sum_{m,n} A_{mn} B_{mn} / \sqrt{\sum_{m_1,n_1} A_{m_1,n_1}^2 \sum_{m_2,n_2} B_{m_2,n_2}^2}$, where $A_{mn}$ and $B_{mn}$ are the intensities of the pixel in the $m_{th}$ row and $n_{th}$ column of the retrieved and original images, respectively[43, 44]. Taking the background into consideration, we define the relative similarity as $S_r = (S - S_{bg})/(1 - S_{bg})$, where $S_{bg}$ is the similarity between the original pattern and the background. We use the relative similarity in this work because $S_r$ approaches zero when the pattern is completely smeared out.

The temperature of the ensemble is measured to be approximately 200 μK in a time-of-flight experiment, and the most probable distance of atomic motion is calculated to be approximately 35 μm after 20 μs of storage. In addition to the thermal motion, a neutral atom with a magnetic moment experiences a force in the gradient magnetic field. When the ambient electric field can be neglected, the neutral atom motion can be calculated as $\dot{p} = -(\mu_F \cdot \nabla) B(r)$. The gradient of the magnetic field $\partial_{x_i} B_j$ is measured to be on the order of $O(\sim 1\ Gs/m)$, and thus, the acceleration of an individual atom is on the order of $O(\sim 10^{-2}\ m\cdot s^{-2})$. During a short storage time (tens of microseconds), the extra drift displacement originating from the magnetic gradient is on the order of $O(\sim 10^{-3}\ nm)$, which is much smaller than that of the atom thermal motion. Therefore, treating atoms as motionless particles in the simulation when the storage time is relatively short (<20 μs) is reasonable since the average motion is much smaller than the size of the Fraunhofer diffraction pattern stored in the ensemble.

## SIMULATIONS AND EXPERIMENTAL RESULTS

First, we store the same pattern with three different sizes, and the experimental results are shown in Fig. 2 (b). As the pattern size decreases, the overlap between diffracted patterns in the FP and the inhomogeneous magnetic field increases; therefore, the distortion effect of the inhomogeneous magnetic field on the stored pattern becomes more apparent.

From the theory, we find that the gradient of the inhomogeneous magnetic field, instead of the magnetic field itself, causes the deformation of the stored pattern. When a uniform polarization magnetic field much stronger than the inhomogeneous magnetic field is applied (the strength of the polarization field is shown in Fig. 3), the total ambient field in the vicinity of the ensemble can be seen as a uniform field in the transverse plane, $\partial_x B = \partial_y B = 0$, and hence, the precession of atomic magnetic moments is synchronized. Therefore, $\eta$ and $\phi$ are also uniform in the transverse plane. Although the destructive interference of different spin-wave components cannot be eliminated in this situation, the transverse distribution of the DSPs in the ensemble is preserved. As shown in Fig. 3 (a), the pattern configuration remains even when the retrieval efficiency is very low while the intensity of the retrieved signal shows Larmor precession. This synchronization effect can dramatically increase the relative similarity of the retrieved pattern (see Fig. 3 (b)). According to the theory, the polarization magnetic field is not necessarily parallel to the quantization axis. We achieved similar results by applying a polarization magnetic field perpendicular to the quantization axis (see Fig. 3 (c)).

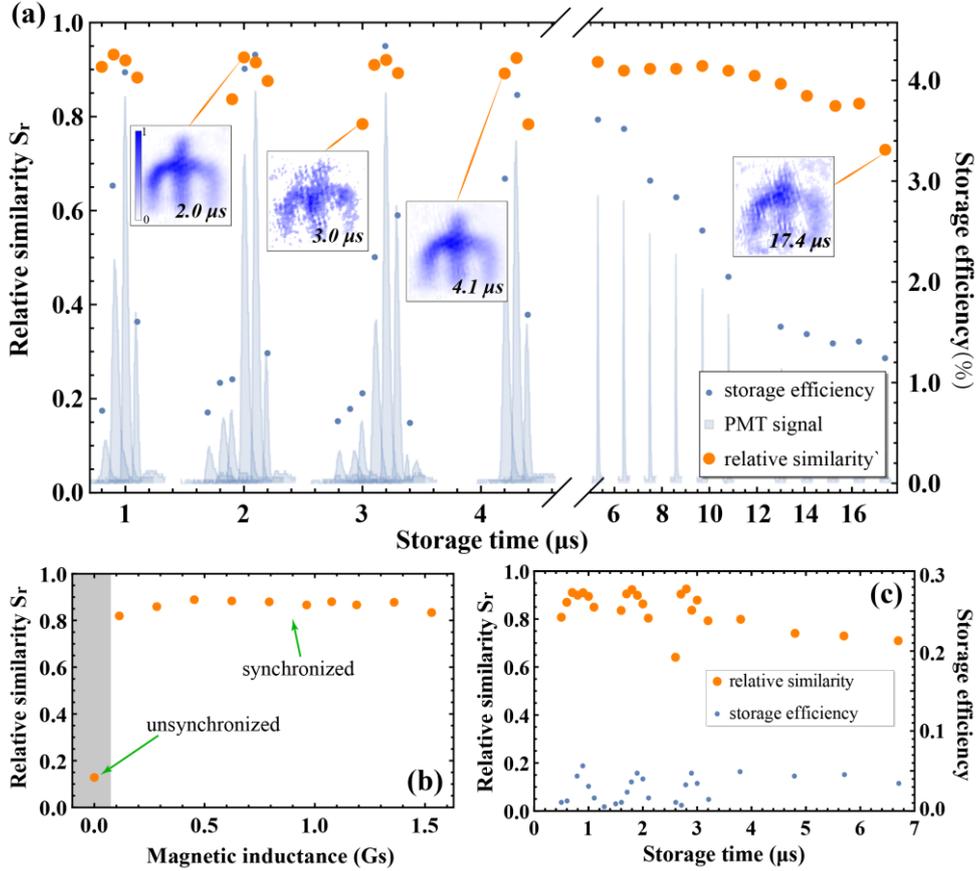

**Fig. 3 (a)** Evolution of DSPs and relative similarity with storage time. A static uniform polarization magnetic field ($B_P$≈0.97 Gs) much larger than the inhomogeneous magnetic field ($B_i$≈0.05 Gs) and parallel to the quantization axis is applied, and Larmor precession appears while the pattern of the retrieved signal remains. (b) When keeping the storage time unchanged ($t_s$=6 μs) and increasing the strength of the polarization magnetic field, as long as the magnetic field is strong enough to synchronize the atomic magnetic moments, the relative similarity increases dramatically. (c) When the direction of the polarization magnetic field is chosen to be perpendicular to the quantization axis, a similar result as in Fig. 3 (a) can be achieved.

In contrast to the former case, when the homogenous polarization magnetic field is switched off, the retrieved pattern is dramatically blurred, as shown in Fig. 4. Because the motion of atoms can be neglected and the static uniform magnetic field has been compensated, this deformation phenomenon mainly originates from the inhomogeneous magnetic field.

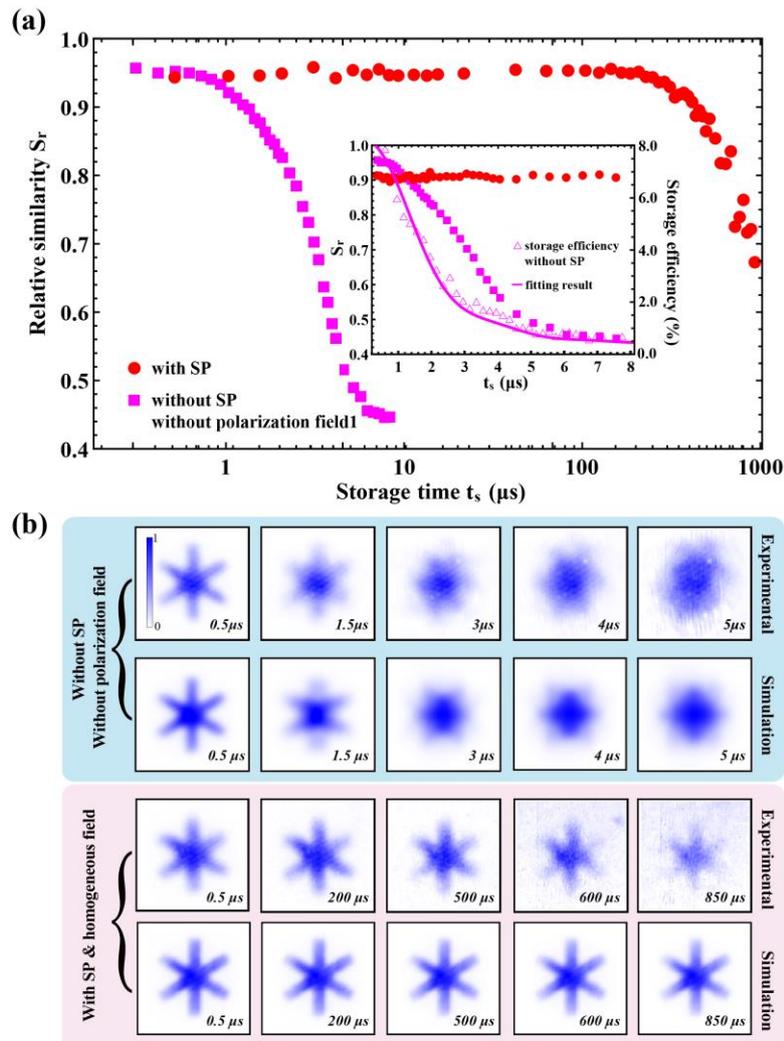

**Fig. 4** (a) Relative similarity and storage efficiency as a function of storage time with SP and without SP. The storage time of the pattern is extended by two orders of magnitude after SP. The inset shows the decay of $S_r$ and storage efficiency during a short storage time; the solid magenta line is the fitting result; and the fitting parameters are utilized in the simulation. (b) Retrieved pattern recorded by the ICCD and corresponding simulation results with and without SP. The pattern could be preserved for up to 800 μs after SP. The simulation result in the last row has taken atomic diffusion into consideration using the theory proposed in ref[30, 32].

To achieve long-lived storage for arbitrary transverse multimodes, one needs to eliminate the decoherence of DSPs. Here, we prepare the initial states as magnetically insensitive states of Rb85, and the SP process is detailed in the Appendix section. After SP, the atoms in the ensemble are in |g, $m_F=0$>. When the signal light and coupling light are tailored to be orthogonally linearly polarized, only one group of spin waves $S_{s,0}^{g,0}$ exists. Hence, the destructive interference among different groups of spin waves disappears; therefore, the decoherence of DSPs is dramatically reduced. Moreover, because the polarization magnetic field is chosen to be parallel to the quantization axis, we have $D_{m,0}^F = \delta_{m,0}$ for arbitrary hyperfine level $F$ and quantized sublevel $m$; therefore, only an overall phase factor is picked up in the transverse plane during the precession, and the retrieved pattern remains unchanged. The above process can be schematically illustrated in an intuitive picture, depicted as case C in Fig. 1 (b): all the atomic magnetic moments are perpendicular to the quantization axis, and the projection of the atomic magnetic moment on the quantization axis remains zero during the precession. The experimental result of the relative similarity as a function of storage time is plotted in Fig. 4 (a), and the retrieval efficiency is also shown in Fig. 4 (a). The solid magenta line is the fitting result of the retrieval efficiency, and the fitting parameter is used as the strength of the inhomogeneous magnetic field in the simulation of Fig. 4 (b). As shown in Fig. 4 (b), we achieve an obvious prolongation of pattern preservation with time after SP. In the third row, the spatial information of the pattern is preserved for up to 800 μs after SP, which is prolonged by two orders of magnitude compared to the first row, where the retrieved pattern is dramatically blurred after 2 μs of storage. The simulation results shown in the second and fourth rows agree with the experimental results in the first and third rows. The simulation results in the fourth row have taken atomic diffusion into consideration, and we use the theory proposed in ref[31, 33]. Moreover, we find that after SP, the difference between the decay rates of the relative similarity of different-size patterns decreases compared to Fig. 2 (b) because the deteriorating effect of the inhomogeneous magnetic field has been diminished.

To illustrate that the deformation of the retrieved pattern is dominated by the inhomogeneous magnetic field, we turn off the homogeneous polarization field after the spin waves are established, and the majority of spin waves are in $S_{s,0}^{g,0}$; see the time sequence shown in Fig. 5 (a). After 95 μs of storage, we turn on a small inhomogeneous magnetic field, and the average intensity of this field is calculated to be on the order of O(~1*10$^{-2}$ Gs), which is large enough to change the configuration of the existing residual homogeneous polarization field (the average intensity is calculated to be on the order of O(~1*10$^{-2}$ Gs)). We found that even though the majority of spin waves are initially in $S_{s,0}^{g,0}$ after the SP process (the SP efficiency is measured to be approximately 70% in an independent experiment), the unsynchronized precession of atomic magnetic moments in the inhomogeneous ambient magnetic field still leads to deformation of the retrieved pattern. The relative similarity as a function of storage time is shown in Fig. 5 (b). $S_r$ obviously decays faster in an inhomogeneous field. The corresponding CCD signal is shown in Fig. 5 (c). The retrieved pattern remains nearly unchanged in a homogeneous ambient field, while the transverse information is smeared out after the inhomogeneous ambient field is turned on for 10 μs.

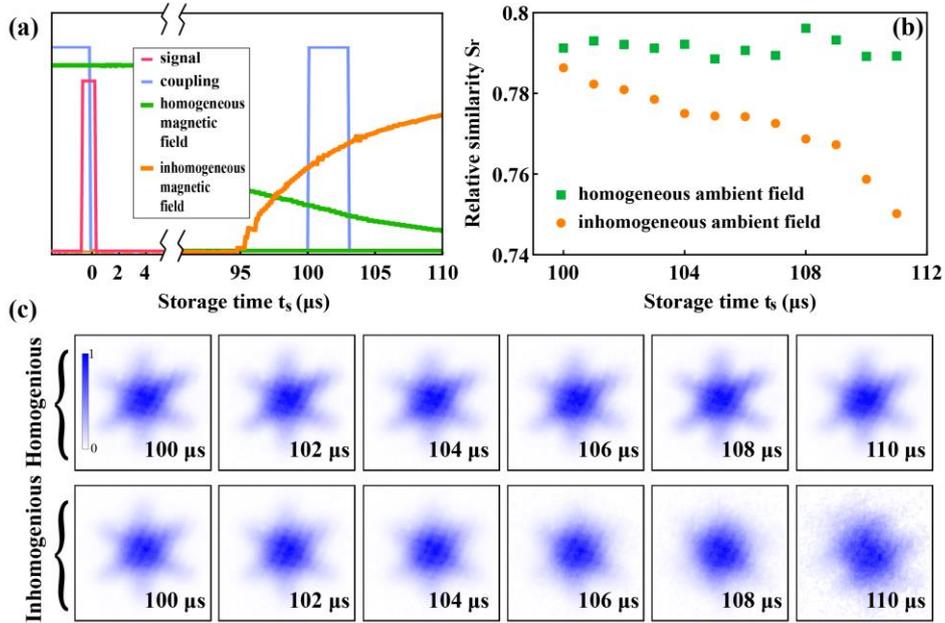

**Fig. 5** (a) Time sequence of the experiment. (b) Relative similarity as a function of storage time with and without an inhomogeneous ambient magnetic field. The inhomogeneous magnetic field is applied at $t_s$=95 μs. (c) Retrieved pattern recorded by the CCD.

## CONCLUSIONS

In conclusion, we theoretically and experimentally demonstrate that the inhomogeneous magnetic field is the main distortion effect on the stored image. The stored transverse multimode with a wider spatial frequency distribution in the FP is more vulnerable to the deformation effect because it has a larger overlap with the inhomogeneous magnetic field. The simulation results based on the model show good agreement with the experimental data. According to the model, we propose and demonstrate that a uniform polarization magnetic field can eliminate this deformation effect by synchronizing the Larmor precession. Additionally, we illustrate that by preparing the ensemble in magnetically insensitive states, the decoherence of DSPs and dephasing of the stored transverse multimode can be diminished, and the storage time is prolonged by two orders of magnitude. The approach we proposed demands neither a passive magnetic shield made of a high-permeability soft magnetic material nor an active compensation method that requires fast detection of the ambient magnetic field in a MOT. Our work may benefit future research on the long-lived storage of transverse multimodes using technologies such as dipole trapping or dynamic decoupling[45, 46] and has potential applications in quantum communications.

## APPENDIX

**Absorption cell filtering.** Although PBS2, PBS3 and the Glan-Tylor prism work as polarization filters for orthogonally polarized signal and coupling light, a more delicate filter is needed to filter the strong coupling light out when the signal light and coupling light are propagating collinearly. The traditional Fabry–Pérot etalon is not suitable because not every transverse multimode can survive in the optical cavity. Therefore, we use the Rb85 absorption cell as a narrow bandpass filter. As shown in the inset of Fig. 1 (c), expanded pump light passes through the absorption cell in a direction nearly opposite to that of the signal light. The frequency of the pump light is detuned from the optical transition $|g, m_F=0\rangle \leftrightarrow |e, m_F=0\rangle$ by an AOM to reduce the influence of the Doppler effect.

**Time sequence.** Our experiment runs periodically with a repetition rate of 50 Hz. In each cycle, the cooling light is turned off 500 μs before the repump light is turned off so that all the atoms can be prepared in the |g> state. The polarization coil is turned on 60 μs before the MOT starts to be turned off to establish the SP magnetic field. The experimental window opens 300 μs after the MOT starts to be turned off. Before the experimental window opens, two SP light beams are turned on for 50 μs. The exposure time of the ICCD is set to 100 ns, which matches the width of the retrieved pulse, and every captured retrieved image presented in this article is the average result of 1000 captures.

**State preparation (SP).** We use a two-frequency optical pumping setup[47, 48]. The relative energy levels are shown in Fig. 1 (a). By illuminating the ensemble with π light (SP II in Fig. 1 (a), 1 mW) that is resonant with the |F=2>→|F'=2> transition and linearly polarized light (SP I in Fig. 1 (a), 1 mW, resonant with the |F=3>→|F'=2> transition) propagating along the quantization axis simultaneously, the population of atoms in |g, $m_F=0$> in the ensemble accumulates because the optical transition $\Delta F = 0, m_F = 0 \rightarrow m_{F'} = 0$ is dipole forbidden; after a period of time, the majority of atoms (approximately 70%) are in the desired state. The advantage of this two-frequency optical pumping method is that it does not require the creation of a population difference between different Zeeman sublevels or hyperfine states. The polarization magnetic field should be large enough to split the EIT transmission peak[49]; after SP, only one transmission peak dominates.

**Inhomogeneous magnetic field.** We place two pairs of coils around the MOT (see Fig. 1 (c)). Coil pair I, which is parallel to the signal light, can generate a strong uniform polarization magnetic field in the vicinity of the ensemble and works as polarized coils. In the experiment, the applied inhomogeneous magnetic field is from a pair of anti-Helmholtz coils (coil pair II). To probe the inhomogeneous magnetic field generated by coil pair II, we use another coil pair III (not shown in Fig. 1 (c)) as probe coils, which has an inductance much smaller than that of the polarized coils and coil pair II and is perpendicular to coil pair I. By observing the voltage across the probe coils during the experiment, we can detect the inhomogeneous magnetic field. We ignore the mutual inductance between the probe coils and other coils; thus, the rate of change in the inhomogeneous magnetic field strength in the vicinity of the ensemble can be estimated by $d\bar{B}(t)/dt \approx -\chi \mathrm{U}(t)/(\mathrm{NS})$. Here, N and S are the number of turns and the cross-sectional area of the probe coils, respectively. The coefficient $\chi$ is the ratio of the average intensities of the magnetic field near the ensemble and the probe coils, which is calculated to be 0.024 in our experiment. The change in the inhomogeneous magnetic field during 20 μs of storage is estimated to be on the order of O($\sim 1*10^{-4}$ Gs), which is much smaller than its initial strength; therefore, this inhomogeneous magnetic field can be seen as static during the simulation.